\documentclass[a4paper]{jpconf}
\usepackage{amssymb}
\usepackage{amsfonts}
\usepackage{amsmath}
\usepackage{bbm}
\usepackage{graphicx}
\usepackage{color}
\usepackage{rotating}
\usepackage{lscape}
\usepackage{lscape}
\usepackage{wasysym}
\usepackage{hyperref}
\usepackage{mathtools}
\usepackage{float}
\def\Z{\mathbb Z}

\def\NN{\mathbb N}

\renewcommand{\th}{\theta}

\newcommand{\im}{{\mathrm{i}}}

\newcommand{\+}{{\dagger}}
\newcommand{\sfrac}[2]{{\textstyle\frac{#1}{#2}}}
\newcommand{\half}{{\sfrac12}}
\newcommand{\pa}{\partial}

\newcommand{\ph}{\phantom{-}}

\newcommand{\beq}{\begin{equation}}
\newcommand{\eeq}{\end{equation}}
\newcommand{\eq}{\end{equation}}
\newcommand{\bea}{\begin{eqnarray}}
\newcommand{\eea}{\end{eqnarray}}

\newcommand{\for}{{\quad{\rm for}\quad}}
\renewcommand{\and}{{\quad{\rm and}\quad}}

\renewcommand{\=}{\ =\ }

\begin{document}

\title{\mbox{Algebraic
integrability of ${\cal PT}${-}deformed Calogero models}}

\author{Francisco Correa$^1$ and Olaf Lechtenfeld$^2$ \\${}$}

\address{$^1$Instituto de Ciencias F\'isicas y Matem\'aticas,\\
Universidad Austral de Chile, Casilla 567, Valdivia, Chile\\
${}^{}$}

\address{$^2$Institut f\"ur Theoretische Physik and Riemann Center for Geometry and Physics\\
Leibniz Universit\"at Hannover \\
Appelstra\ss{}e 2, 30167 Hannover, Germany}

\ead{francisco.correa@uach.cl, olaf.lechtenfeld@itp.uni-hannover.de}

\begin{abstract}
We review some recents developments of the algebraic structures and spectral properties of non-Hermitian deformations of  Calogero models. The behavior of such extensions is illustrated by the $A_2$ trigonometric and the $D_3$  angular Calogero models. Features like intertwining operators and conserved charges are discussed in terms of Dunkl operators. Hidden symmetries coming from the so-called algebraic integrability for integral values of the coupling are addressed together with a physical regularization of their action on the states by virtue of  a $\mathcal{ PT}$-symmetry deformation.
\end{abstract}

\section{Introduction}
\vskip 0.5cm
Calogero models, also known as Calogero--Moser--Sutherland models, represent one of the best examples of many-particle integrable models and find applications in a wide range of areas in physics and mathematics. Introduced first by Calogero for  pairwise inverse-square interactions with three and $n$ particles \cite{Cal69}, it was then generalized to different type of potentials. The rational potential can be extended to a trigonometric, a hyperbolic \cite{suther} or an elliptic one. Moreover, all mentioned cases can be formulated for any finite Coxeter group \cite{OlshaPere81-rev, OlshaPere83-rev}, enabling a large class of many-particle integrable models.  There is a vast literature on this topic; for an overview in the subject and many of the applications, see for instance \cite{OlshaPere81-rev, OlshaPere83-rev,Poly06-rev,bmodels, cmsbook}. In recent decades,  the familiy of Calogero systems has been studied under the light of non-Hermitian Hamiltonians \cite{ali,rev,qscs}. Non-Hermitian extensions include a wide range of integrable systems, see for instance \cite{andreasreview}, and realizations in nature due to the application of integrable non-linear equations in optics \cite{bookpt}.  In this work, we focus in two particular features of Calogero models: ${\cal PT}$-symmetric deformations and the algebraic structure related with conserved quantities and intertwining operators. Both topics are briefly reviewed here. \newpage

The first non-Hermitian extensions of  Calogero models were done back in 2000 by Znojil and Tater, performing imaginary shifts on the coordinates in two- and three-particle systems \cite{znotat}.  In the same year Basu--Mallick and Kundu proposed an extension of the rational $A_{n-1}$ Calogero model inspired by long-range interaction with explicit momentum dependence \cite{BasuMallick:2000af}. Despite their model not being Hermitian, the energy eigenvalues are real and bounded from below.  This idea was extended later
in the ${\cal PT}$-symmetric regime to other models \cite{ BasuMallick:2001ce}, see also \cite{BasuMallick:2002kg, BasuMallick:2003pt, BasuMallick:2004ye}.  A next step further was done by Fring in \cite{Fring:2005ys}, studying the extensions in \cite{BasuMallick:2000af, BasuMallick:2001ce, BasuMallick:2003pt, BasuMallick:2004ye} from a generic perspective including all Coxeter groups beyond the rational case, towards the trigonometric, hyperbolic and elliptic models. As a result, the rational non-Hermitian deformations turned out to remain integrable,  but all other cases require compensating terms to keep integrability.  Until now, the most elegant way to introduce non-hermiticity is enlarging the Coxeter root systems \cite{frzn, assisandreas}. Ways to generically construct  complex root systems were developed in a series of papers by Fring and Smith \cite{fringsmith1,fringsmith2,fringsmith3}. Other results for  Calogero models in the non-Hermitian realm include, analysis of  complex domains \cite{vitcalogero}, quasi exactly solvable approaches \cite{Brihaye:2003dc}, complex extensions of the coupling constants \cite{Ghosh:2003gz}, random matrix theories \cite{randomm, jain}, spectral singularities \cite{Mandal:2012ww}, isospectral  and supersymmetric deformations \cite{Ghosh:2010ic,Ghosh:2011tu}. More recently,  ${\cal PT}$-symmetric deformations of Calogero models have been used to construct invisible and reflectionless potentials by means of complex Darboux transformations related with the Korteweg--de Vries integrable hierarchy \cite{cjp} playing a role in conformal and supersymmetric theories \cite{MateosGuilarte:2017fsv, MateosGuilarte:2018gxu,Inzunza:2021vgt}. The quantum behavior of Calogero systems from a Hamiltonian formulation considering balanced gain and loss was studied in  \cite{Ghosh:2017rgp,Sinha:2017lck}, for a recent review see \cite{Ghosh:2021ucx}. The idea of introducing non-Hermiticity was also studied from the point of view of spectral degeneracies, conserved quantities and intertwining operators \cite{CoLe17, Correa:2019hnu}. The objective of this brief review is to discuss the main results of these works and the future prospects for the topic. \newline

Intertwining operators for Calogero models were introduced in the 1990s \cite{ChaVes90,Chalykh96} connecting the Liouville integrals at different coupling values. They play a crucial role when the couplings take integer values, allowing one to obtain Liouville eigenstates from the free theory but also to build up algebraically independent conserved quantities, on top of the Liouville integrals and beyond superintegrability. In this regime the models are known to be algebraically or analytically integrable.  All those features can be adressed by means of Dunkl operators \cite{Dunkl89}, see also \cite{Opdam,Heckman}. These integrals were treated as conserved quantities in formal sense since they commute with the Liouville integrals. In the rational case, they generate supersymmetric algebras \cite{clp, Carrillo-Morales:2021ugo}. However, the action of the additional charges is not well defined, mapping physical states to non-physical ones. We show that this can be remedied by means of a ${\cal PT}$-regularization. In fact, the idea of healing the action of the additional conserved charges is not new and has been studied in one dimensional cases \cite{cp1,cp2} but also in regularizing degenerate soliton solutions of the Korteweg--de Vries equation \cite{Correa:2016zrb}. \newline


This paper is organized as follows. In Section 2 we summarize the main features of the trigonometric
Calogero--Sutherland model for the $A_2$ root system. Both conserved quantities and intertwining operators are presented together with their algebra and their action on the energy eigenstates. Then we introduce ${\cal PT}$-symmetry in a simple way in order to discuss the spectral degeneration and the physical restoration of a nonlinear conserved charge. A similar approach is given in Section 3 but for the angular Calogero model associated with the $D_3$ root system. The last section is devoted to conclusions and open problems.

\section{${\cal PT}$-symmetry in Calogero--Sutherland models}
\vskip 0.5cm
The quantum Calogero--Sutherland Hamiltonian \cite{suther} was introduced as a toy model in nuclear physics due to the type of short-range interaction in comparison with the rational version. We consider three interacting particles with coordinates $x_i \in {\mathbb R}/2\pi{\mathbb Z}, \, i =1,2,3$, on a circle governed by the $A_2$ Coxeter root system and the Hamiltonian
\begin{equation}\label{hcs}
H_{CS}(g)=-\frac{1}{2}\sum^3_{i=1} \pa_i^2
+\sum_{i<j}^3\frac{g(g{-}1)}{\sin^2 (x_i{-}x_j)} \ ,
\end{equation}
where $g$ is a coupling parameter. As we shall see below this coupling parameter plays an important role in relation with degeneracy and conserved quantities in the non-Hermitian case. The spectral and algebraic properties of the system (\ref{hcs}) can be studied by different methods,  in the  present discussion we focus on the approaches given in \cite{Poly1,LaVi95,PeRaZa98,GaLoPe01}. In particular, we will use the Dunkl operator approach, which in this model takes the form
\begin{equation}\label{dunkl}
D_i(g) \= \pa_i - g\sum_{j(\neq i)} \cot (x_i{-}x_j)\,s_{ij}
\end{equation}
where  $s_{ij}$ permutes the coordinates $x_i$ and $x_j$. This method is particularly useful for a number of reasons, including the construction of
\begin{itemize}
\item all conserved quantities by means of Weyl-\emph{invariant} polynomials in the $D_i$ operators,
\item intertwining operators  by means of Weyl-\emph{anti-invariant} polynomials in the $D_i$ operators,
\item the energy eigenstates in terms of the Jack polynomials in an algebraic manner.
\end{itemize}
These features will be briefly revisited below. For the $A_2$ root system described by (\ref{hcs}), the conserved quantities are constructed by means of the Newton sums
\begin{equation}
{I}_m(g)=\textrm{res} \left[D_1^m(g)+D_2^m(g)+D_3^m(g)\right] \ , \quad m=1,2,3 \ ,
\end{equation}
but we have only three independent integrals.  The notation ``\textrm{res}" stands for the restriction to completely symmetric functions, which removes all permutation operators. The selection the charges is not unique, any other permutation-invariant polynomial in the Dunkl operators will also provide an integral of motion. Instead,  the following basis is considered, 
\begin{equation}\label{good}
C_1(g) = I_1(g) \ ,\quad C_2 = I_2(g)  -8g^2 = -2\,H_{CS}(g) ,\quad C_3(g)  = I_3(g)  - I_1(g)  I_2(g)  \ ,
\end{equation}
which besides satisfying $[C_i(g) ,C_j(g) ]=0$ provide the simplest intertwining relations.  The intertwining operators are constructed by the restriction of any permutation anti-symmetric polynomial. The simplest one has differential order three,
\begin{equation}\label{inter}
M(g) \= \sfrac13\,\textrm{res}\,\bigl( 
D_{12}(g)D_{23}(g)D_{31}(g)+D_{23}(g)D_{31}(g)D_{12}(g)+D_{31}(g)D_{12}(g)D_{23}(g) \bigr) \ ,
\end{equation}
where we denote $D_{ij}(g)=D_i(g){-}D_j(g)$.  Further higher-order intertwining operators from anti-symmetric polynomials may be constructed following the same recipe. In the current case only one intertwiner is needed for a complete algebraic description. However, as we shall see in the next section, sometimes higher-order ones are required.  The explicit expressions of conserved quantities (\ref{good}) and intertwiners (\ref{inter}) are given in \cite{Correa:2019hnu}. With the basis (\ref{good}), the intertwining relations take the standard form
\begin{equation}\label{intercs}
M(g)\,C_\ell(g) \= C_\ell(g{+}1)\,M(g)\ .
\end{equation}
As a consequence of the above relation, the action of intertwining operators will not change the energy on the wavefunctions. In fact, (\ref{intercs}) is nothing else than the shape-invariant feature studied in the context of supersymmetric quantum mechanics. The presence of shape-invariance and the construction of the spectrum have been studied in Calogero models \cite{Efthimiou:1996qw,Ghosh:1997yy} but using first order interwiners of a different nature. The energy spectrum of the stationary Schr\"odinger equation
\begin{equation}\label{hcseq}
H_{CS}(g)\Psi_{n_1,n_2}^{(g)}=E_{n_1,n_2}\Psi_{n_1,n_2}^{(g)} \ ,
\end{equation}
depends quadratically on two quantum numbers, $n_1$ and $n_2$, 
\begin{equation}\label{ener}
E_{n_1,n_2}(g) = (n_1+2g)^2 + \sfrac13(n_1-2n_2)^2,
\end{equation}
obeying the relation $n_1\ge n_2 \ge 0$ \cite{Correa:2019hnu,LaVi95}. The wavefunctions 
\begin{equation}\label{eigen}
\Psi_{n_1,n_2}^{(g)}\=\,e^{-\sfrac{2\im}{3}(n_1+n_2)(x_1+x_2+x_3)}\,\Delta^g  P_{n_1,n_2}^{(g)}(x_1,x_2,x_3)
\end{equation}
are given in terms of the Vandermonde determinant $\Delta=\prod_{i<j} \sin (x_i{-}x_j)$ and the so-called Jack polynomials $P_{n_1,n_2}^{(g)}$, which are homogeneous polynomials of degree $n_1+n_2$ in the $x_i$ coordinates and  symmetric under permutations.  They can be constructed analytically  in terms of  deformed Dunkl operators. For more details of their construction and properties, see \cite{LaVi95, Correa:2019hnu} and references therein. The action of the intertwining operators on the wavefunctions reads
\begin{align}\label{theshift}
M(g)\,\Psi_{n_1,n_2}^{(g)} &\= n_2 (n_1{+}g) (n_1{-}n_2) \Psi_{n_1-2,n_2-1}^{(g+1)} \ ,\\
M^\dagger(g)\,\Psi_{n_1,n_2}^{(g)} &\= (n_1{+}3g{-}1) (n_1{-}n_2{+}2g{-}1) (n_2{+}2g{-}1) \Psi_{n_1+2,n_2+1}^{(g-1)} \ ,
\end{align}
where $M(1{-}g)=M^\dagger(g)$. As the action on the intertwining operators on the wavefunctions does not change the energy values (\ref{intercs}), the shifting on the $g$ parameter in (\ref{ener}) is compensated with modifications in the quantum numbers $n_1$ and $n_2$. The conserved quantities (\ref{good}) action on the states take the form
\begin{align}
C_1(g) \Psi_{n_1,n_2}^{(g)} &=0\, ,\\
 C_2(g) \Psi_{n_1,n_2}^{(g)}& =2\bigl[(n_1{+}2g)^2+\sfrac13(n_1{-}2n_2)^2\bigr]\Psi_{n_1,n_2}^{(g)}\,  ,\\ \label{c3}
C_3(g) \Psi_{n_1,n_2}^{(g)}&=-\sfrac{8}{9}\im\,(n_1{-}2n_2)(2n_1{-}n_2{+}3g)(n_1{+}n_2{+}3g)\Psi_{n_1,n_2}^{(g)}\ .
\end{align}
In the second relation, (\ref{ener}) is used into the definition of $C_2$ in coherence with the changes in (\ref{theshift}). The degeneration due to $n_2\mapsto n_1{-}n_2$ flips the overall sign of the action of $C_3$ in (\ref{c3}), and therefore these two degenerate energy states can be distinguished by this integral of motion.
\vskip 0.5 cm

\emph{Issues on the algebraic integrability and the symmetry restoration by ${\cal PT}$  deformations}
\label{2p1}
\vskip 0.5 cm
The idea of revisiting the non-Hermiticity in the Calogero--Sutherland model is inspired mainly by three problems.
\begin{enumerate}
\item 
In the Hermitian case discussed above, the ground state of (\ref{hcs}) is given by 
\begin{equation}
\Psi_{0,0}^{(g)}=\prod_{i<j}^3 \sin^g (x_i{-}x_j)\ , \quad \text{with} \quad 
E_{0,0}(g)=4g^2 \ .
\end{equation}
It is clear what $\Psi_{0,0}^{(g)}$ vanishes when the coordinate values coincide, those regions correspond to the Weyl-alcove walls defining also the singularities of the interacting potential (\ref{hcs}) given by the $A_2$ structure.  Because of the power dependence, when $g<0$ the ground state and more generically the wave-functions $\Psi_{n_1,n_2}^{(g)}$ become non-physical due to the non-normalizability resulting from such singularities. 
\item The Calogero--Sutherland Hamiltonian (\ref{hcs}) displays a naive but relevant symmetry changing the coupling constant parameter according to $g\leftrightarrow1{-}g$. This symmetry suggests that states of two different values can be considered simultaneously within a single unique Hilbert space. However, because of the previous point, states containing negative powers of $g$ will be non-physical making this symmetry meaningless.
\item The cases when $g\in\NN$ are special. In this situation the Calogero--Sutherland model is called ``algebraically integrable". This feature appears by a combination of the symmetry $g\leftrightarrow1{-}g$ and the intertwining operator $M(g)$. As the action of a single interwiner shifts the coupling constant by unity, for integer values of the coupling, we can step from $1-g$ to $g$ and vice versa by iteration of the process. In other words, we can build a chain of $2g{-}1$ consecutive interwiners in the form
\begin{equation}
Q(g) \= M(g{-}1) M(g{-}2) \cdots M(2{-}g) M(1{-}g)
\end{equation}
which acting on the Hamiltonian and using the invariance under  $g\leftrightarrow1{-}g$, gives
\begin{align}
Q(g) H(g)&=H(1{-}g) Q(g)=H(g) Q(g) \ .
\end{align}
In this way the operator $Q(g)$ turns out to be an extra conserved quantity. However, because of the previous arguments for $g>1$, the action of $Q(g)$ will transform physical states into non-physical ones, and the reverse for $g<0$.
 \end{enumerate}
 \vskip 0.25 cm
 
\begin{figure}[H]
\centering
\includegraphics[scale=0.4]{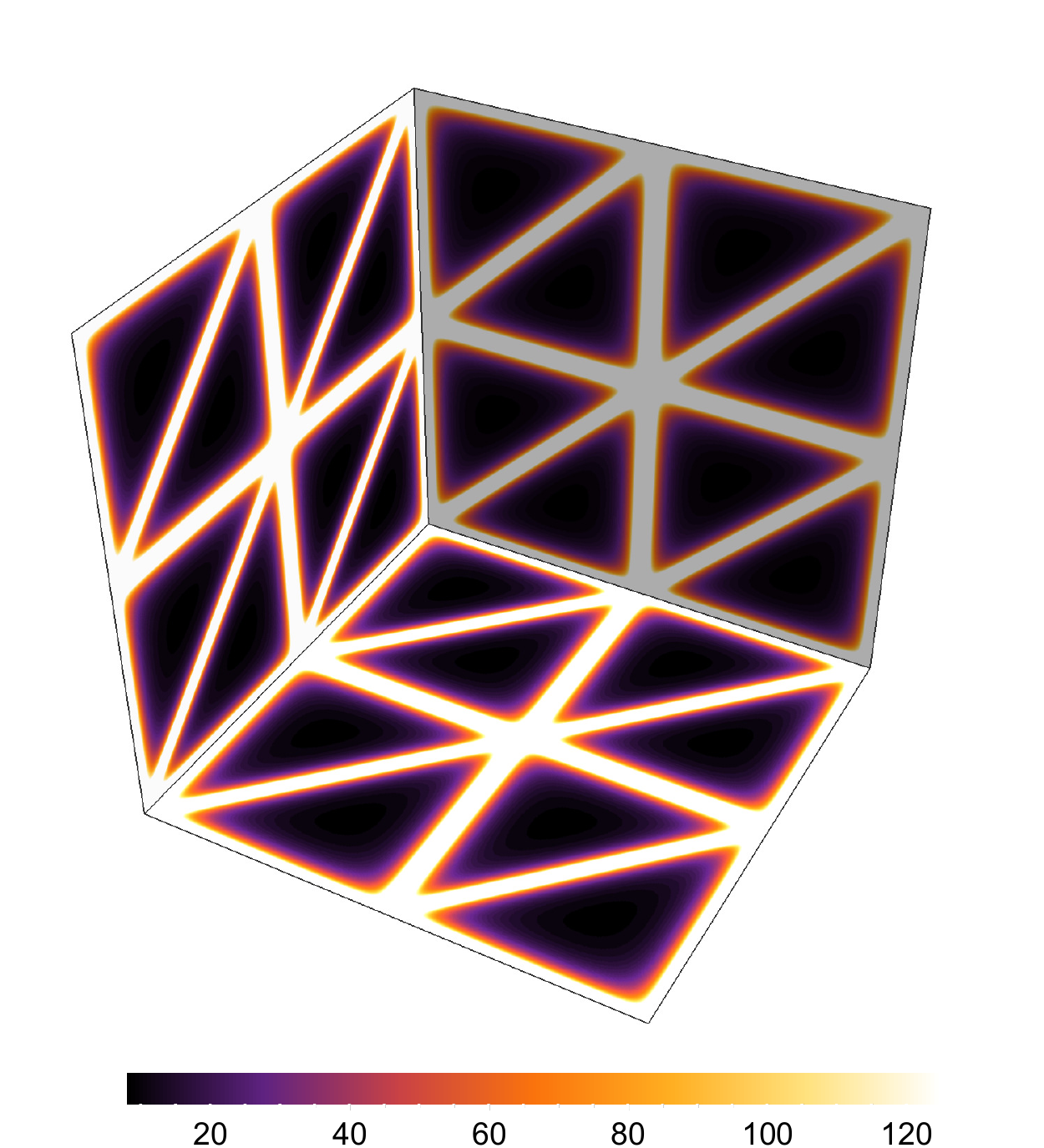}
\includegraphics[scale=0.4]{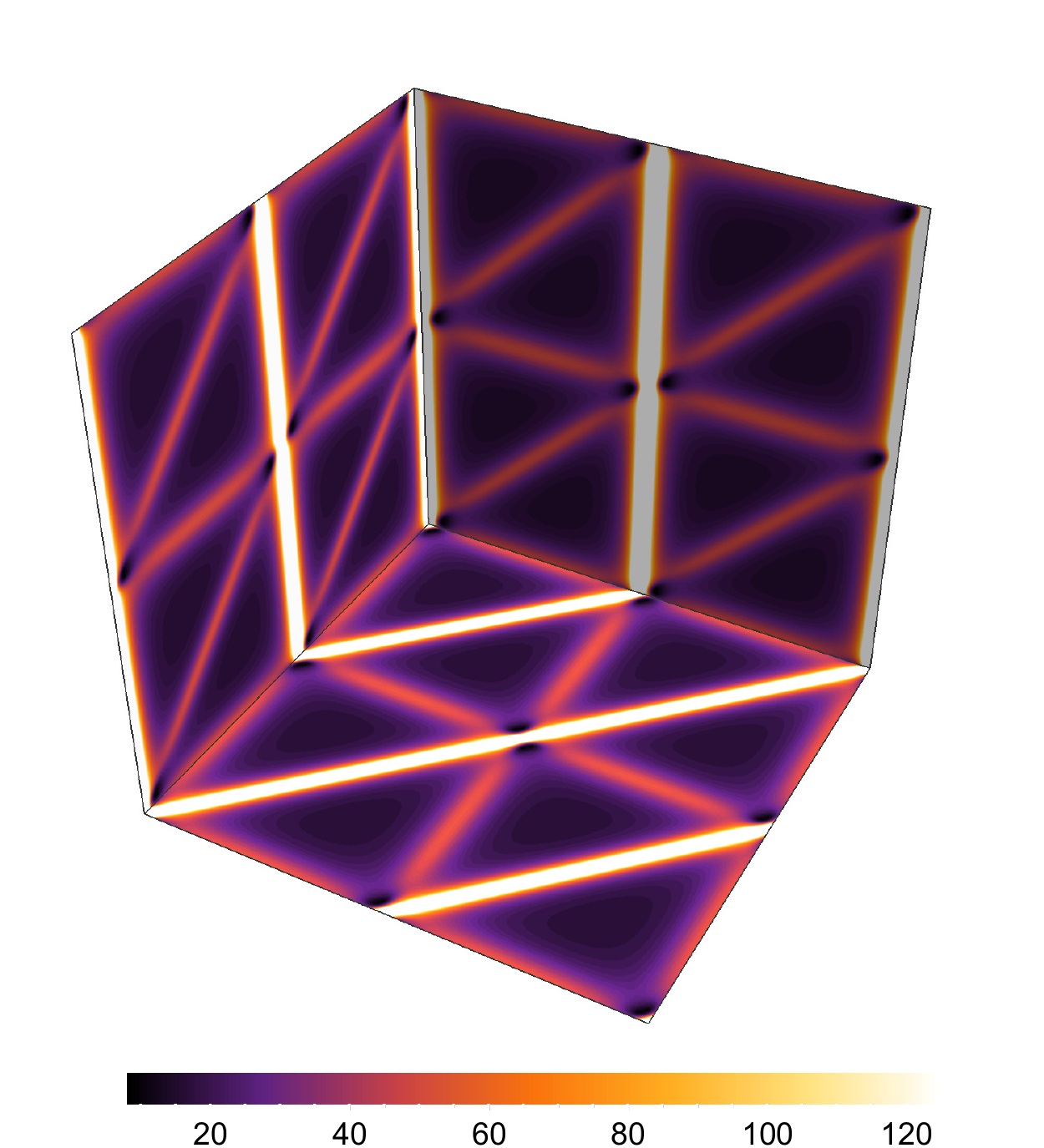}
\includegraphics[scale=0.4]{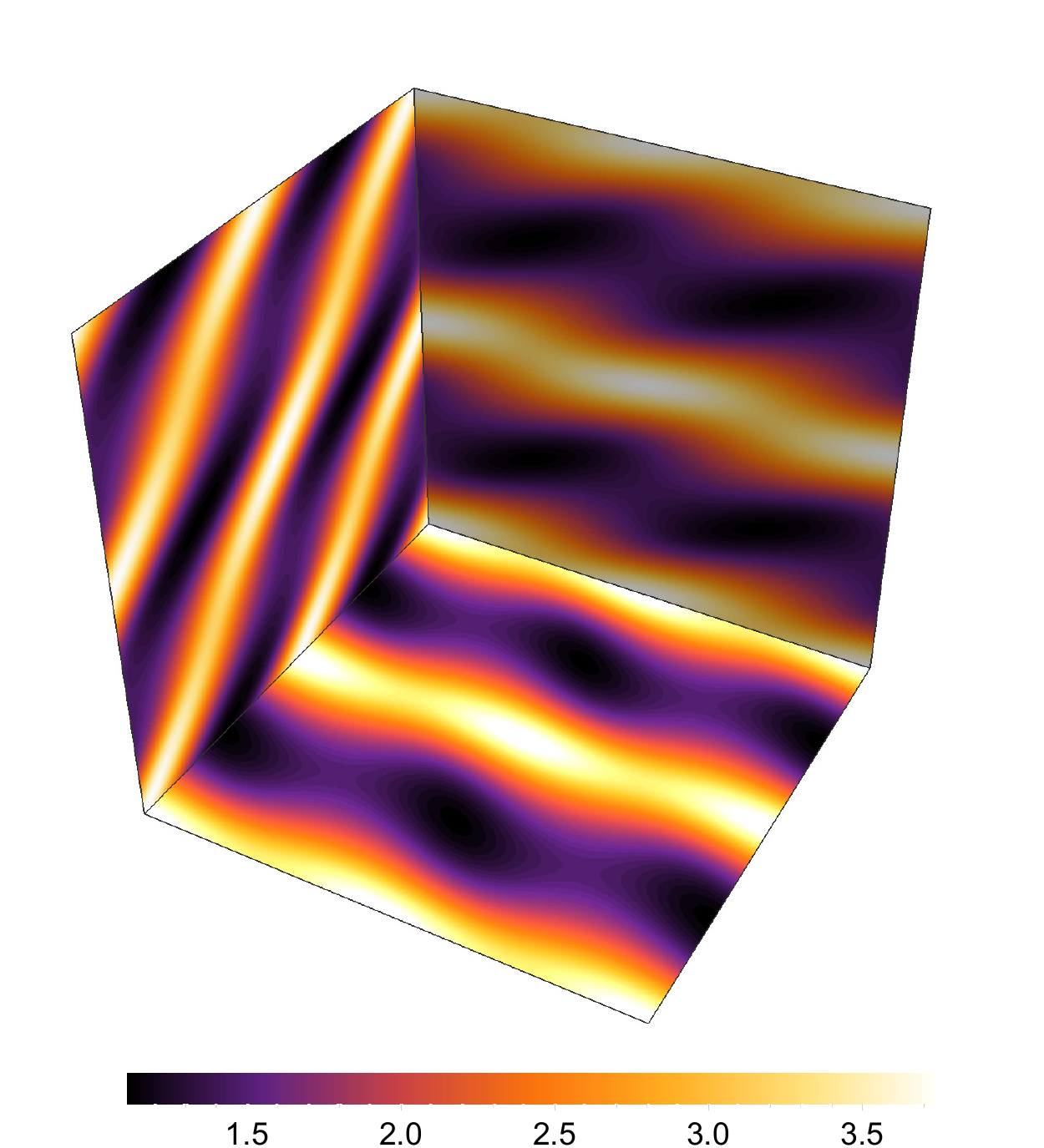}
\caption{3D sliced density plots the absolute value of the potential term in (\ref{hcs}). (Left) The pure real potential with all $\epsilon_\ell=0$. The boundaries of the Weyl alcoves appear as white lines. (Center) The same plot but with $\epsilon_3=0.2$ turned on. The potential still displays some singular regions. (Right) When all parameters are turned on  $\epsilon_1=2, \epsilon_2=-0.5$ and $\epsilon_3=0.2$,  the potential is regularized. Note the change of the scale in comparison with other plots.
 }
\label{fig1}
\end{figure}

We can tackle all these points at the same time by introducing ${ \cal PT}$-symmetry into the system. Among the different approaches one may follow \cite{znotat, Fring:2005ys, frzn,fringsmith1,fringsmith2,fringsmith3,Correa:2019hnu}, here we will use the simplest one by shifting the coordinates by an imaginary amount,
\begin{equation}\label{reg}
x_\ell \rightarrow x_\ell +\im \epsilon_\ell, \quad \ell=1,2,3.
\end{equation}
In this way the consider both ${\cal P}$ and ${\cal T}$ operators in a standard way,
\begin{equation}
{\mathcal P} : \ (x_1,x_2,x_3)\mapsto(-x_1,-x_2,-x_3), \quad \text{and} \quad{\mathcal T} : \im \mapsto - \im \ .
\end{equation}
In order to find a complete regularization of the system we must turn on all three parameters $\epsilon_\ell$. Figure~\ref{fig1} shows how the absolute value of the potential looks when the complex regularization (\ref{reg}) is introduced.

\begin{figure}[h!]
\centering
\includegraphics[scale=0.7]{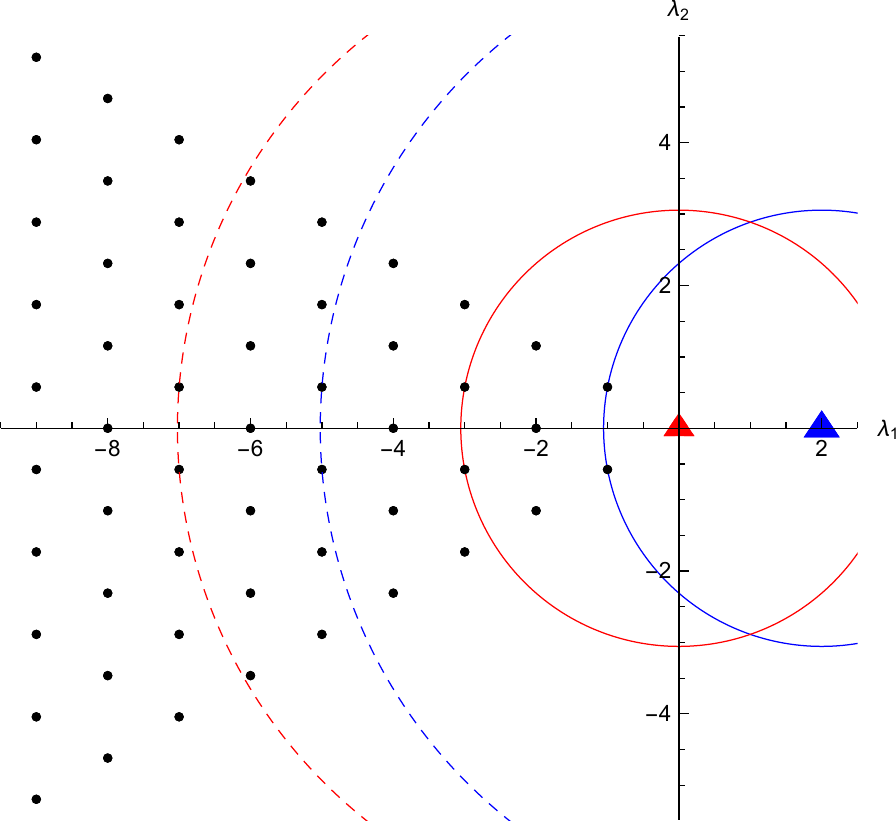}
\qquad
\includegraphics[scale=0.82]{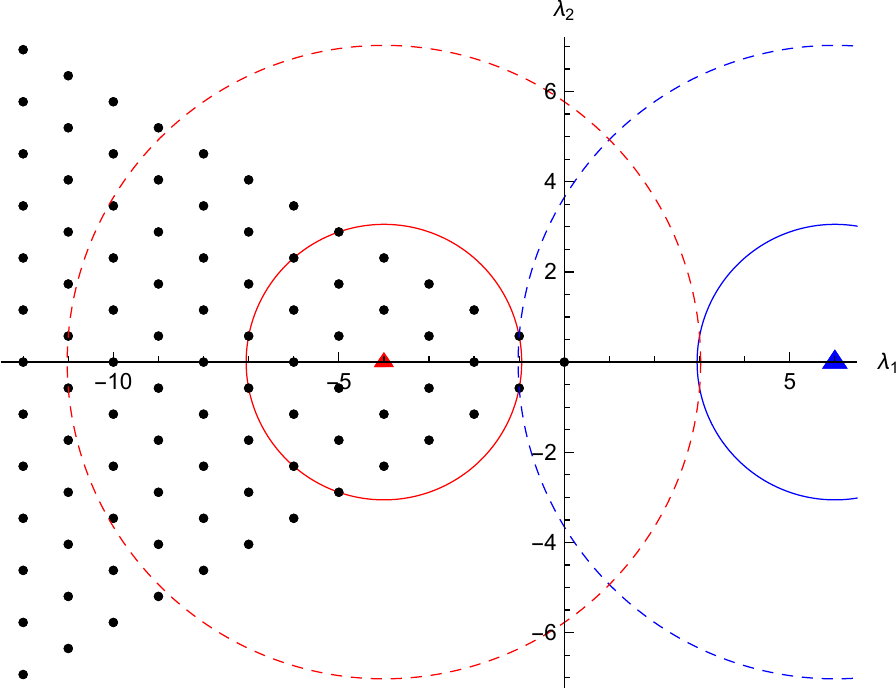}
\caption{In the weight space, the states are represented as black dots and  circles correspond to fixed ``energy shells''. (Left) Solid and dashed lines stand for energies $E=\sfrac{28}{3}$ and $E=\sfrac{148}{3}$ while blue and red colors stand for $g{=}1$ and $g{=}0$ respectively. (Right) Same energy shells as before but blue and red colors correspond to $g{=}3$ and $g{=}{-}2$ respectively.} 
\label{fig2}
\end{figure}

Once the potential does no longer display singularities we are able to use the symmetry $g\leftrightarrow1{-}g$ and join the states from both sides considering $g>\sfrac{1}{2}$. The enhancement of energy degeneracy becomes apparent when we write the energy (\ref{ener}) in the weight space notation
\begin{equation}
E_{n_1,n_2}(g) = (\lambda_1-2g)^2+\lambda_2^2\ .
\end{equation}
Here we identify
\begin{equation}\label{lsphere}
(\lambda_1,\lambda_2) = \bigl(-n_1,\sfrac{1}{\sqrt{3}}(n_1{-}2n_2)\bigr) \ ,
\end{equation}
and the condition $n_1\ge n_2 \ge 0$  is translated into $\lambda_1\le-\sqrt{3}\,|\lambda_2|$. Thus, in the $\lambda$-space the set of all allowed states form a $\sfrac{\pi}{3}$ wedge, as can be seen in Figure~\ref{fig2}. Considering a circle centered at $(2g,0)$ of radius $R_g=\sqrt{E_{n_1,n_2}(g)}$, all the states lying on the circle will share the same energies. For instance, the states with $\pm \lambda_2$, i.e.  $\Psi_{n_1,n_2}^{(g)}$ and $\Psi_{n_1,n_1-n_2}^{(g)}$, belong to those cases. On top of that, after  ${\cal PT}$ regularization and because of the  symmetry $g\leftrightarrow1{-}g$, we can also take into account the states on the circle centered at  $2(1{-}g,0)$ of same radius. Albeit the cases with $g\geq 0$ are rarely degenerated, the cases when  $g< 0$ display a high degeneracy, up to order 12.  

\begin{figure}[h!]
\centering
\includegraphics[scale=0.8]{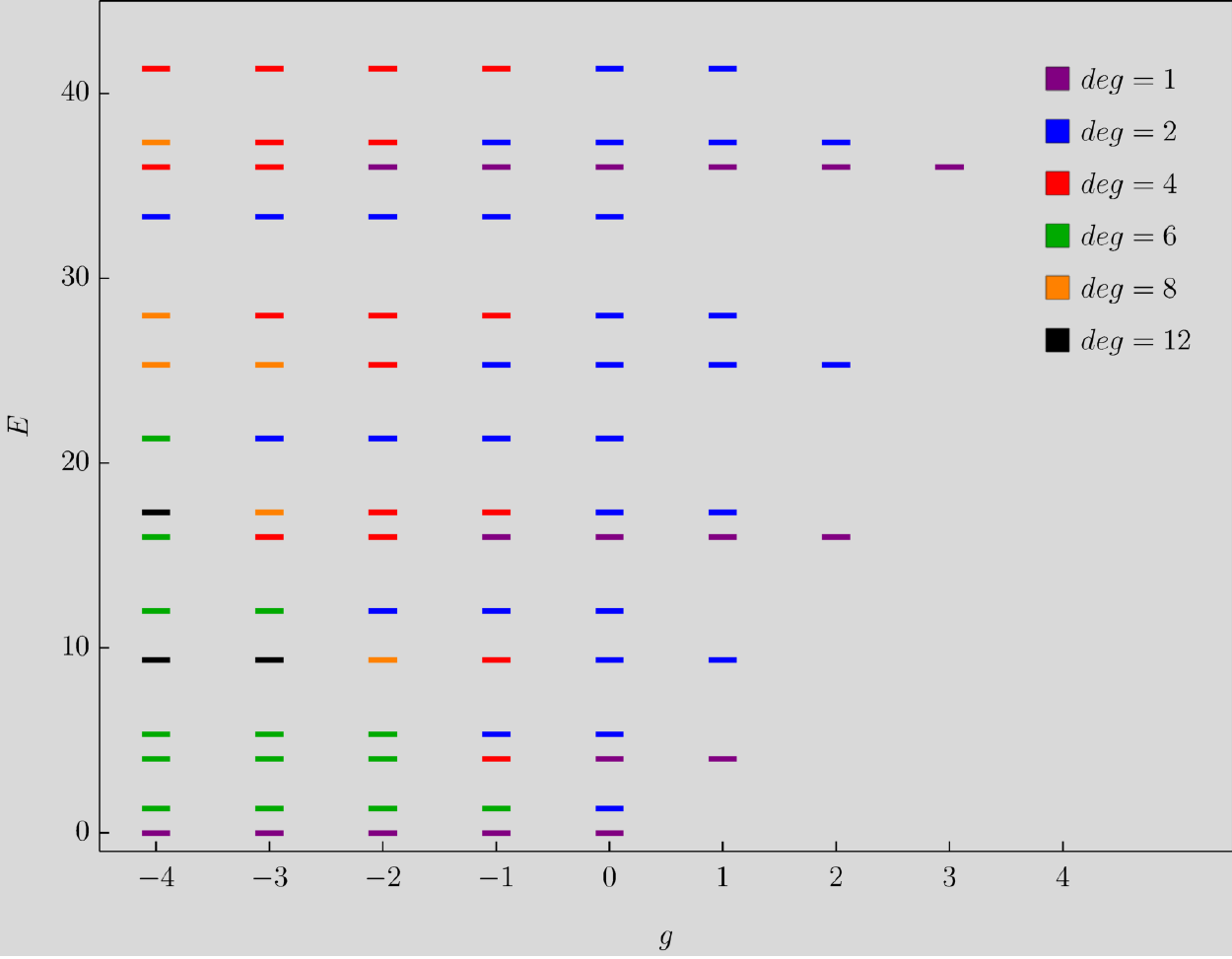}
\caption{Energy spectrum and degeneracies for the $A_2$ model at integer coupling $g\in\Z$. }
\label{fig3}
\end{figure}

As we mentioned above, the case  $g\in\Z$ is peculiar, and the  degeneracy related with $g\leftrightarrow1{-}g$ is also reflected
by an extra conserved charge $Q(g)$, which has always odd differential order $3(2g{-}1)$. If we define that states with  $g{>}0$ and $g{\le}0$ have certain parity, for instance even and odd respectively, 
then the operator $Q(g)$ is of odd nature,  in the sense that it maps those sets of states into each other. The explicit action on the states (\ref{eigen}) takes the form
\begin{equation}\label{eqq}
Q(g)\Psi_{n_1,n_2}^{(g)} \propto \Psi_{n_1-4g+2,n_2-2g+1}^{(1-g)} \ ,
\end{equation}
 and we note that both states belong to the same Hilbert space due to $g\leftrightarrow1{-}g$. In fact, $Q(g)$ does not change the Liouville eigenvalues because it commutes with all charges,
 \begin{equation}
[Q(g),C_\ell(g)]=0, \quad \ell=1,2,3.
\end{equation}
The idea of making well-defined the action of an odd type of conserved charge is not new and has been studied
in one-dimensional cases in the past \cite{cp1,cp2}. The notion of $Q(g)$ as an odd integral, allows one to build different types of hidden supersymmetry structures without fermion degrees of freedom \cite{Plyushchay:1994re, Plyushchay:1999qz}. In Calogero models we can choose as a grading operator any of the permutations $s_{ij}$  due to $\{Q(g),s_{ij}\}=0$, and therefore $Q(g)$ may be treated as a supercharge. This notion of algebraic structures was studied in detail for the rational Hermitian Calogero model \cite{clp}, see also \cite{Carrillo-Morales:2021ugo}. It is natural to wonder whether the operators $Q(g)$ are completely independent of the Liouville integrals. After shifting the coupling from $g$ to $g+1$ by the action of $M(g)$, one may shift it back to $g$ by applying $M (1{-}g)=M^\dagger (g)$. Therefore
the combination $M^\dagger (g) M(g)$ should commutes with the Hamiltonian. We can verify this by virtue of 
\begin{equation}\label{mm}
M^\dagger (g) M(g)=R(g) = 18C_3^2+8C_3C_1^3-3C_2^3+3C_2^2C_1^2-C_2C_1^4+C_1^6 -6g^2(3C_2-C_1^2+8g^2)^2 \ ,
\end{equation}
which is nothing else than a polynomial in the conserved charges. We can elucidate the meaning of $Q(g)$  by taking its square,
\begin{align}\notag
Q^2(g)&= M(g{-}1) \cdots M(3{-}g)  M(2{-}g) M(1{-}g)M(g{-}1) M(g{-}2) M(g{-}3)\cdots  M(1{-}g)  \\ \notag
&= M(g{-}1) \cdots M(3{-}g) M(2{-}g) M^\dagger(g{-}1)M(g{-}1) M(g{-}2) M(g{-}3)\cdots  M(1{-}g)  \\ \notag
&= M(g{-}1) \cdots M(3{-}g)M(2{-}g) R (g{-}1)M(g{-}2) M(g{-}3)\cdots  M(1{-}g)  \\
&= M(g{-}1)  \cdots M(3{-}g)M(2{-}g) M(g{-}2) R (g{-}2)M(g{-}3)\cdots  M(1{-}g)  \\ \notag
&= M(g{-}1)  \cdots M(3{-}g)\left(R (g{-}2)\right)^2M(g{-}3)\cdots  M(1{-}g)  \\ \notag
& \vdots \\ \notag
&=M(g{-}1)M(1{-}g)  \left(R (1{-}g)\right)^{2g-2} =\left(R (1{-}g)\right)^{2g-1} =(M^\dagger (g) M(g))^{2g-1} \ .
\end{align}
So using the relation (\ref{mm}) we identify $Q^2(g)$ as a higher-order polynomial in the conserved charges. Thus $Q(g)$ is not a standard supercharge but a nonlinear supercharge in a wider sense, see \cite{clp}.
We finish this section presenting  in Fig.~\ref{fig3}  a plot of the different degeneracies for the low values of $g$.

\newpage

\section{${\cal PT}$-symmetry in angular Calogero models}
\vskip 0.5 cm

\begin{figure}[h!]
\centering
\includegraphics[scale=0.53]{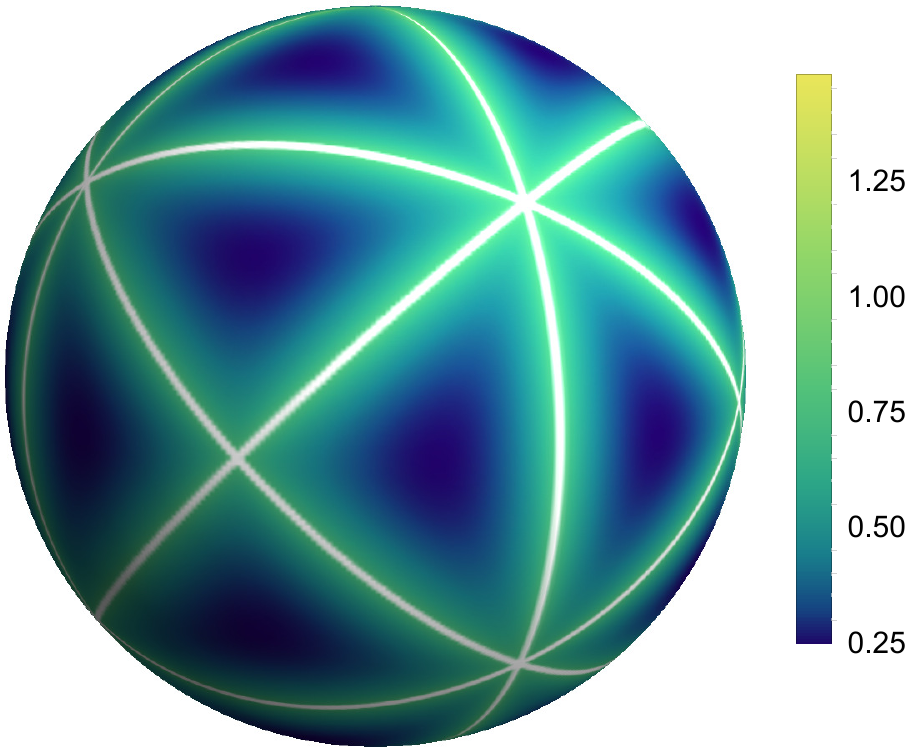}\qquad
\includegraphics[scale=0.53]{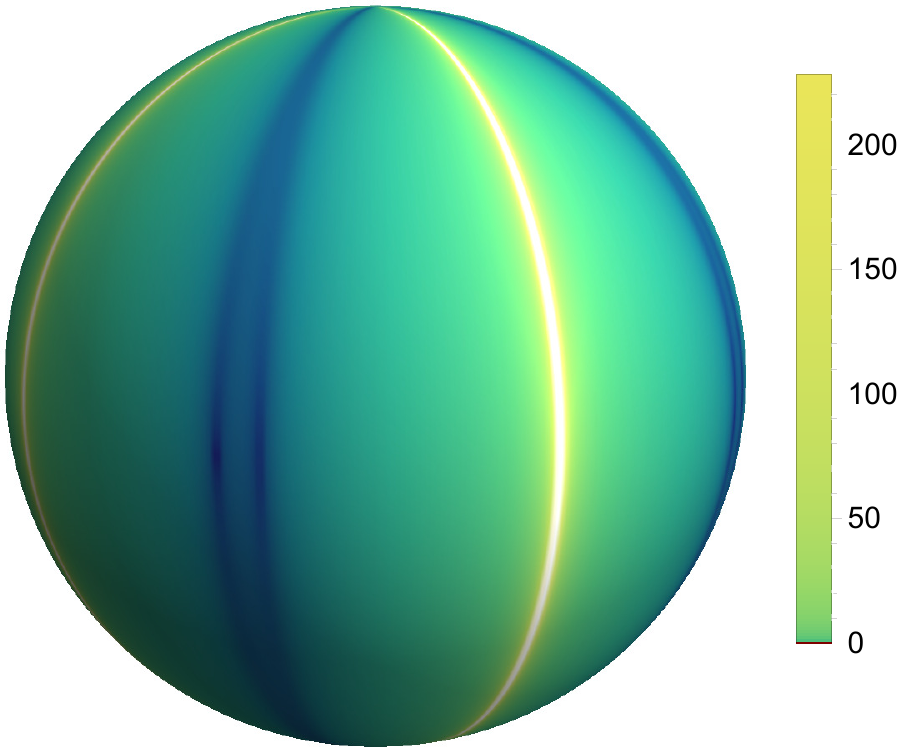}\qquad
\includegraphics[scale=0.53]{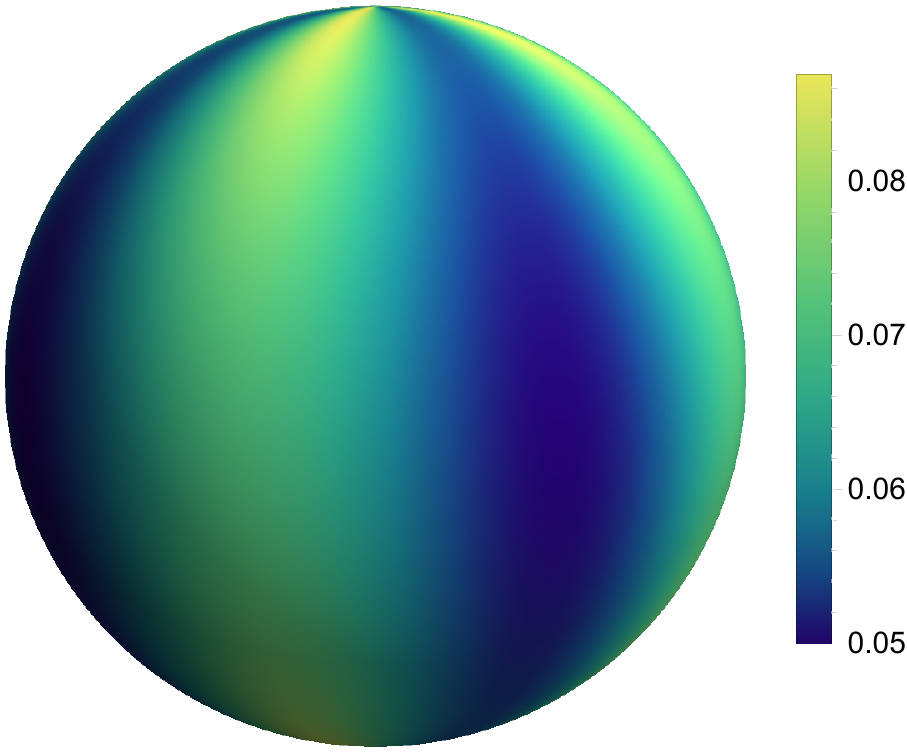}
\caption{Density plots of the potential term in (\ref{angularh}) before and after the ${\cal PT}$ regularization, see (\ref{regu}) below. (Left) In the Hermitian case, $\epsilon_1=\epsilon_2=0$,  the plot is scaled by a composition of a term $\log \circ \log\circ \log$. The Weyl walls are represented by the white lines. (Center) Absolute value of the potential in the non-Hermitian case with $\epsilon_1=2.5$ and $\epsilon_2=0$, where some singular lines remain present. (Right) Absolute value of the complete regularized potential for $\epsilon_1=2.1$ and $\epsilon_2=1.5$.}
\label{fig4}
\end{figure}

In this section we review a different kind of many-particle system and its non-Hermitian extension. The $n$-particle Calogero model with rational interaction potential displays a conformal symmetry, which enables the superintegrability of the system \cite{Woj, Kuz}. Alternatively, the system may be interpreted as a conformal particle living in ${\mathbb R}^n$ and being subject to an external potential. It is possible to separate the radial and the angular part of the Hamiltonian which defines an angular Calogero model living on the hypersphere $S^{n-1}$. One may naively guess that the angular system is simpler than the original one but, despite the fact the angular model is still superintegrable, the converse is true. This is why the angular Calogero models have been studied recently at the classical and quantum level \cite{CoLe17,HNY,HKLN,HLNS,HLN,FeLePo13,CoLe15}, but also regarding some of their algebraic structures \cite{Feigin03,FeHa14,FeHa119}. In order to illustrate the main features of these models, we focus on the $D_3$ angular version. A more detailed discussion of the following ideas is given in \cite{CoLe17,CoLe15}. The Hamiltonian in this case includes the angular momentum as a kinetic term and a tetrahexahedral potential,
\begin{equation}\label{angularh}
 H(g)=-\frac{1}{2}\sum_{i<j}^3(x_i\pa_j{-}x_j\pa_i)^2 +2\,g(g{-}1)\,(x_1^2{+}x_2^2{+} x_3^2)\, \sum_{i<j}^3 \frac{x_i^2{+}x_j^2}{(x_i^2{-}x_j^2)^2}\ ,
\end{equation}
and is also invariant under  $g\leftrightarrow1{-}g$. Figure~\ref{fig4} shows the potential and the tessellation of the sphere in 24 isosceles triangles defined by the Weyl chambers.

In the same spirit of the previous section, we focus the discussion on the Dunkl operator approach. The analogues of (\ref{dunkl}) involve the angular momenta instead of the linear ones and take the form 
\begin{align}
{\mathcal L}_1&= x_2\pa_3{-}x_3\pa_2 + g\left(
\sfrac{x_3}{x_1-x_2}s_{12}{-}\sfrac{x_3}{x_1+x_2}\widetilde{s}_{12}{-}\sfrac{x_2+x_3}{x_2-x_3}s_{23}{+}
\sfrac{x_2-x_3}{x_2+x_3}\widetilde{s}_{23}{+}\sfrac{x_2}{x_3-x_1}s_{31}+\sfrac{x_2}{x_3+x_1}\widetilde{s}_{31} \right) \ ,\\
{\mathcal L}_2&= x_3\pa_1{-}x_1\pa_3 + g\left(
\sfrac{x_1}{x_2-x_3}s_{23}{-}\sfrac{x_1}{x_2+x_3}\widetilde{s}_{23}{-}\sfrac{x_3+x_1}{x_3-x_1}s_{31}{+}
\sfrac{x_3-x_1}{x_3+x_1}\widetilde{s}_{31}{+}\sfrac{x_3}{x_1-x_2}s_{12}+\sfrac{x_3}{x_1+x_2}\widetilde{s}_{12} \right)\ , \\
{\mathcal L}_3&= x_1\pa_2{-}x_2\pa_1 + g\left(
\sfrac{x_2}{x_3-x_1}s_{31}{-}\sfrac{x_2}{x_3+x_1}\widetilde{s}_{31}{-}\sfrac{x_1+x_2}{x_1-x_2}s_{12}{+}
\sfrac{x_1-x_2}{x_1+x_2}\widetilde{s}_{12}{+}\sfrac{x_1}{x_2-x_3}s_{23}+\sfrac{x_1}{x_2+x_3}\widetilde{s}_{23} \right)\ ,
\end{align}
where from now on we omit the explicit dependence on $g$ in the operators unless necessary.
As the dynamics is governed here by the $D_3$ Coxeter group, besides the permutations $s_{ij}$ we introduce the reflections $\widetilde{s}_{ij}$ given by
\begin{align}
\widetilde{s}_{12} :& \ (x_1,x_2,x_3) \mapsto (-x_2,-x_1,+x_3) \ ,   \\
\widetilde{s}_{31} :& \ (x_1,x_2,x_3) \mapsto (-x_3,+x_2,-x_1) \ ,\\
\widetilde{s}_{23} : &\ (x_1,x_2,x_3) \mapsto (+x_1,-x_3,-x_2) \ .
\end{align}
which together generate Weyl group $S_4$. Now, with the angular Dunkl operators we are able to construct 
\begin{itemize}
\item all conserved quantities by means of Weyl-\emph{invariant} polynomials in the ${\mathcal L}_i$ operators.
\item intertwining operators  by means of Weyl-\emph{anti-invariant} polynomials in the ${\mathcal L}_i$  operators.
\item the energy eigenstates in terms of harmonic polynomials.
\end{itemize}
One possible choice to build up the conserved charges is 
\begin{equation}
J_k=\ \textrm{res}\bigl(\mathcal{L}_1^k + \mathcal{L}_2^k + \mathcal{L}_3^k\bigr)
\quad\for k=2,4,6\ ,
\end{equation}
where $J_2 = -2\, H(g) +6g(6g{+}1)$ is the shifted Hamiltonian. The higher-order integrals commute with the Hamiltonian, $[J_2,J_\ell]=0$ for $\ell=4,6$, but, in contrast with the previous case $[J_4,J_6]$ is different from zero so it is not a Liouville system. For the sake of simplicity, we are using Cartesian coordinates to describe the Dunkl operators and the wavefunctions. Nevertheless, the Hamiltonian (\ref{angularh}) is two-dimensional and can be expressed completely in terms of a polar and an azimuthal angle \cite{CoLe17,CoLe15}.  As we have three $(2\times2{-}1)$ integrals of motion,  the system is superintegrable. The  conserved quantities $J_4$ and  $J_6$ have differential order greater than two, hence the two-dimensional system is not separable \cite{mil13}. The peculiarity of the angular model is  revealed by the specific form of the $J_k$ algebra. The non-vanishing commutator reads
\begin{align}
[J_6,J_4] &=\ 12 M_3^\+ M_6 + 24(3{+}4g)J_6J_2 -12(3{+}2g)J_4^2- 48(1{+}2g)J_4J_2^2 + 12(1{+}2g)J_2^4 \\
&+\textrm{lower-order terms}\ .
\end{align}
It cannot be expressed only in terms of the $J_k$ basis integrals and depends explicitly on \emph{two} interwining operators, defined next. In the trigonometric case, only one intertwining operator was required to completely describe the algebraic structure. Here we need two intertwiners of differential order three and six respectively,
\begin{align}
 M_3 &= \sfrac16 \textrm{res}\bigl(
\mathcal{L}_1\mathcal{L}_2\mathcal{L}_3 + \mathcal{L}_1\mathcal{L}_3\mathcal{L}_2 + \mathcal{L}_2\mathcal{L}_3\mathcal{L}_1 + 
\mathcal{L}_2\mathcal{L}_1\mathcal{L}_3 + \mathcal{L}_3\mathcal{L}_1\mathcal{L}_2 + \mathcal{L}_3\mathcal{L}_2\mathcal{L}_1 \bigr)\ , \\
M_6 &= \textrm{res}\bigl(
\{\mathcal{L}_1^4,\mathcal{L}_2^2\}-\{\mathcal{L}_2^4,\mathcal{L}_1^2\}+\{\mathcal{L}_2^4,\mathcal{L}_3^2\}
-\{\mathcal{L}_3^4,\mathcal{L}_2^2\}+\{\mathcal{L}_3^4,\mathcal{L}_1^2\}-\{\mathcal{L}_1^4,\mathcal{L}_3^2\}\bigr)\ .
\end{align}
They intertwine the Hamiltonian in the standard way,
\begin{equation}
 M_s (g) H(g)= H(g{+}1) M_s (g) \ ,
\end{equation}
but the generic intertwining relations for the two charges take a more complicated form in comparison to (\ref{intercs}),
\begin{equation}
M_s(g) J_\ell(g) \= \sum_{s',\ell'} \gamma_{s\ell}^{s'\ell'}(g)\,J_{\ell'}(g{+}1)\,M_{s'}(g) \ ,
\end{equation}
where in the sum of the right-hand side could appear more than one interwiner. The functions $\gamma_{s\ell}^{s'\ell'}(g)$ are polynomials in $g$, see \cite{CoLe15}. We  briefly review the energy spectrum for the angular model (\ref{angularh}),
\begin{equation}
 H(g)\ \Psi_{\ell_3,\ell_4}^{(g)} =E_{\ell}\ \Psi_{\ell_3,\ell_4}^{(g)}\ .
\end{equation}
The energy depends on a combination of two quantum numbers $\ell_3$ and $\ell_4$,
\begin{equation} \label{laener}
E_{\ell}=\half q\,(q{+}1) \qquad\textrm{and}\qquad q \= 6g+\ell \= 6g+3\ell_3{+}4\ell_4\ .
\end{equation}
The allowed values $\ell_3,\ell_4=0,1,2...$ lead to degeneracies firstly for $\ell$ and secondly due to the fact $E_\ell$ is quadratic in $\ell $.  The energy eigenfunctions  can be written as
\begin{equation}\label{angularw}
\Psi_{\ell_3,\ell_4}^{(g)}=(x_1{+}x_2{+}x_3)^{-q/2}\Delta^g h^{(g)}_{\ell_3,\ell_4}(x) \ ,
\end{equation}
where the Vandermonde determinant takes the form $\Delta=\prod_{i<j}  (x_i^2{-}x_j^2)$. The $h^{(g)}_{\ell_3,\ell_4}(x)$ are homogenous polynomials of degree $\ell=3\ell_3+4\ell_4$ in the $x_i$ coordinates.  They can be constructed in terms of Dunkl operators and are invariant under the action of the $S_4$ group. For more details of their construction and specific examples, see  \cite{CoLe17,CoLe15}. So far, there are no closed formulas for the action of the conserved quantities $J_4$,  $J_6$ or the intertwiners $M_3$ and $M_6$. Still, the latter act on the wavefunctions (\ref{angularw}) according to
\begin{equation}
M_s(g)\Psi_{\ell_3,\ell_4}^{(g)} \propto \sum_{\ell'=\ell-6} \mu^{s,\ell_3,\ell_4}_{\ell_3',\ell_4'}(g) \Psi_{\ell_3',\ell_4'}^{(g+1)} \ ,\quad s=3,6 \,
\end{equation}
where the $\mu$'s are some polynomials in $g$. As the wavefunctions (\ref{angularw}) contain $\Delta^g$,   for $g<0$ we find singularities at the vanishing locus of the Vandermonde determinant, forcing $g\geq 0$ for a physical spectrum. The degeneracy for the allowed energy levels in this case can be computed exactly and reads
\begin{equation}\label{deg1}
\textrm{deg}(E_\ell) \= \Bigl\lfloor\frac{\ell}{12}\Bigr\rfloor\ +\ 
\begin{cases} 0 & \for \ \ell=1,2,5\ \ \textrm{mod}\ 12 \\
              1 & \for \ \ell=\textrm{else}\ \ \textrm{mod}\ 12 \end{cases}\ .
\end{equation}

\emph{ ${\cal PT}$-symmetry regularization, once again.}
\vskip 0.5 cm

\begin{figure}[h!]\centering
\includegraphics[scale=0.8]{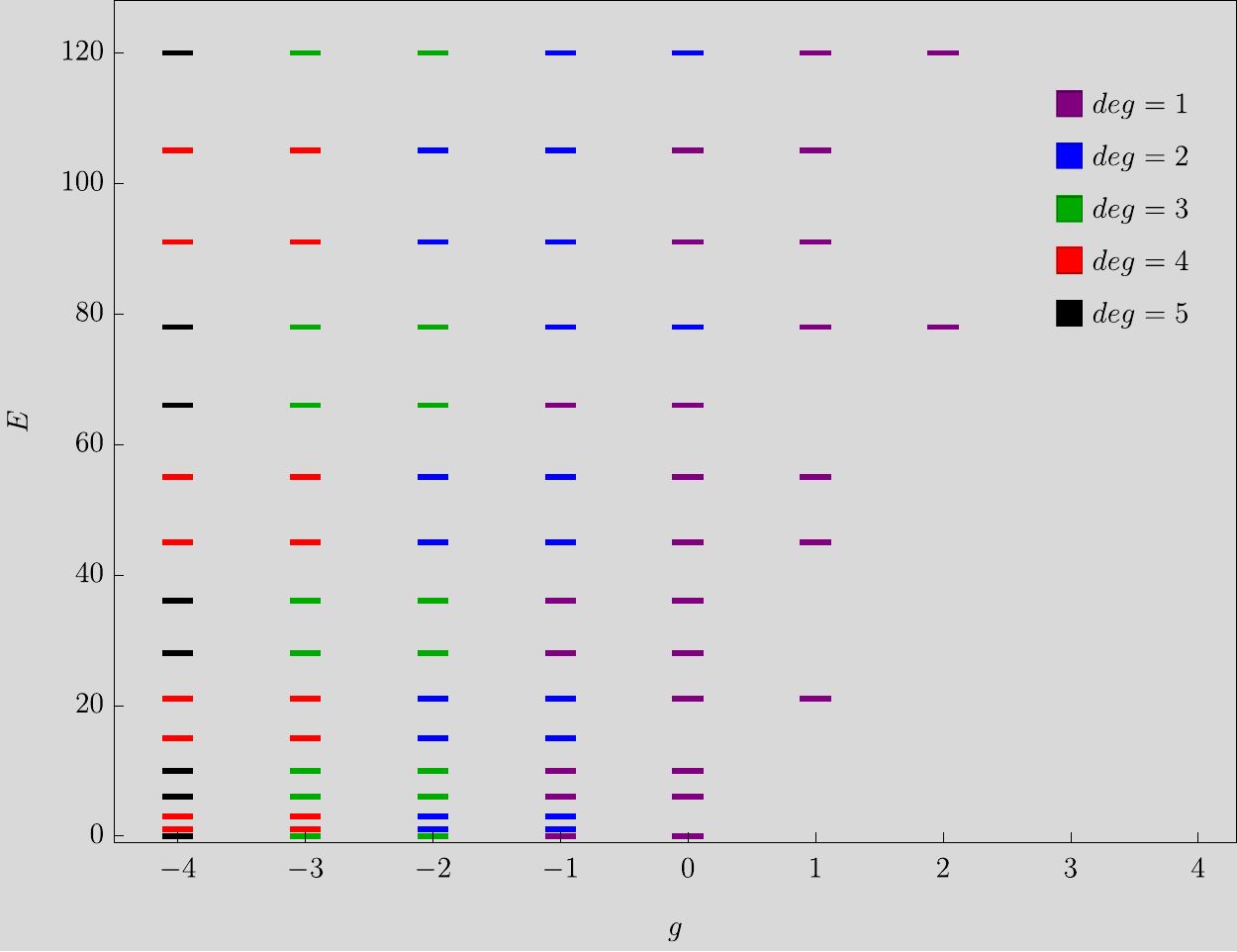}
\caption{Energy spectrum and degeneracies for the $D_3$ model at integer coupling $g\in\Z$.}\label{fig5}
\end{figure}

The set of ideas coming from points (i), (ii) and (iii) in Section~\ref{2p1} also applies to the angular model (\ref{angularh}). It is possible to remove all singularities  by  a ${\cal PT}$-symmetric deformation. This is achieved by introducing spherical coordinates as follows,
\begin{equation}\label{regu}
\begin{pmatrix} x_1\\[1pt] x_2 \\[1pt] x_3 \end{pmatrix} =r \begin{pmatrix} 
\sin(\theta{+}\im\epsilon_1)\cos(\phi{+}\im\epsilon_2) \\[2pt]
\sin(\theta{+}\im\epsilon_1)\sin(\phi{+}\im\epsilon_2) \\[2pt]
\cos(\theta{+}\im\epsilon_1) 
\end{pmatrix}\ .
\end{equation}
The ${\cal PT}$-operator can be chosen as ${\cal P} : \ (\th,\phi)\mapsto(-\th,-\phi)$, which means
\begin{equation}
{\mathcal P} : \ (x_1,x_2,x_3)\mapsto(-x_1,x_2,x_3), \quad \text{and} \quad{\mathcal T} : \im \mapsto - \im \ .
\end{equation}
The Hamiltonian (\ref{angularh}) clearly is invariant under the combined action. In order to remove both potential and wave-function singularities both parameters $\epsilon_1$ and $\epsilon_2$ must be turned on, see Figure~\ref{fig4}. Because of the regularization there now exist physical states for $g<0$, and we must combine them with the tower of states at $1-g>0$. In this way the degeneracy (\ref{deg1}) heavily increases (for large values of the energy) giving as a result
\begin{equation}
\textrm{deg}(E_\ell) \= 
\begin{cases}
\ph g{-}1 \ +\ \begin{cases}
0 &\textrm{for}\quad q+6g=0,3,4,7,8,11\ \textrm{mod} \ 12 \\
1 &\textrm{for}\quad q+6g=1,2,5,6,9,10\ \textrm{mod} \ 12
\end{cases}
\Biggr\} &\textrm{if}\quad q<6g{-}6 \\[4pt]
\ph \Bigl\lfloor\frac{q}{6}\Bigr\rfloor \ \ + \ \begin{cases}
0 &\textrm{for}\quad q=1,2,5\ \textrm{mod} \ 6 \\
1 &\textrm{for}\quad q=0,3,4\ \textrm{mod} \ 6
\end{cases}
\Biggr\} &\textrm{if}\quad q\ge6g{-}6 \, .
\end{cases}
\end{equation}
In Fig.~\ref{fig5} we present the distribution of allowed states and degeneracies for low values of the energy.
 
 \newpage
Like the Calogero--Sutherland model, the angular one becomes analytically integrable for integer values of the coupling constant $g$. However, because we have two different intertwining operators we have more ways to construct the additional odd charges
\begin{equation}
Q(g) = 
M_{*}(g{-}1)M_{*}(g{-}2)\cdots M_{*}(2{-}g)M_{*}(1{-}g)
\end{equation}
where  $M_*$ stands for using in every step either $M_3$ or $M_6$. The odd nature of these conserved charges can understood from the relations 
\begin{equation}
\begin{aligned}
 M_3^\+ M_3^{\vphantom{\+}}  & \propto 
2J_6{-}3 J_4J_2{+} J_2^3+ \ \textrm{lower-order terms} \ , \\
M_6^\+ M_6^{\vphantom{\+}}  &\propto
-12 J_6^2{+}12 \{ J_6, J_4\} J_2{-}\sfrac{16}{3}J_6J_2^3{+}2 J_4^3{-}14 J_4^2J_2^2{+}6J_4J_2^4-\sfrac{2}{3}J_2^6 \\
&{+}\textrm{lower-order terms}\ ,
\end{aligned}
\end{equation}
which tell us that  $Q^2(g)$  is a polynomial in the conserved even charges. For example, in the case of $g=2$ we have
\begin{equation}
\bigl(Q^{(2)}_{333}\bigr)^2=(2J_6{-}3 J_4J_2{+} J_2^3)^3+ \ \textrm{lower-order terms}  \ .
\end{equation}

\section{Outlook and open problems}
\vskip 0.5 cm

In this review we addressed non-Hermitian extensions of the trigonometric and angular Calogero models under the scope of integrability. Both systems exhibit a set of conserved charges, intertwining operators and -for integer  couplings- a higher-order aditional odd integral of motion $Q(g)$. The latter flips the coupling  $g\leftrightarrow1{-}g$ of the states, which means to transform physical states into singular ones. Introducing a ${\cal PT}$-symmetric deformation as a regularization removes all singularities of  potentials and wavefunctions. In this way the conserved charges $Q(g)$ acquire a physical nature. Taking into account the symmetry $g\leftrightarrow1{-}g$,  the spectral degeneracy is radically increased by the deformation. There are further results we have not presented here which are more involved but not less interesting. These features have been studied for the Calogero--Sutherland model $G_2$ model describing the so-called Calogero--Marchioro--Wolfes problem \cite{Qu95}. This is a non-simply-laced case, so there are two couplings associated to the short and long roots of the corresponding Coxeter group, which translates to a richer structure with different types of conserved quantities \cite{Correa:2019hnu}. Regarding the angular model, the $BC_3$, $A_1^{\oplus 3}$ and $H_3$ systems have also been studied in a similar way \cite{CoLe17}. Analogous investigations for the  hyperbolic or elliptic Calogero interactions are still missing. Further deformations of Calogero models  may also be considered \cite{defo}.  For a Hamiltonian discussed there,  can be arranged to
\begin{equation}
H_{D}=-\frac{1}{2}\sum^3_{i=1} \pa_i^2
+\frac{m(m{-}1)}{\sin^2 (x_1{-}x_2)}+\frac{1{-}m}{\sin^2 (x_1{-}i\sqrt{m}  x_3)}+\frac{1{-}m}{\sin^2 (x_2{-}i \sqrt{m} x_3)}, 
\end{equation}
which displays two extra conserved charges. Trigonometric and elliptic deformations of such  systems were also studied, see \cite{new} and references therein. 

\subsection*{Acknowledgments}
\vskip 0.5 cm
FC was partially supported by Fondecyt grant 1211356. OL has been supported by the Deutsche Forschungsgemeinschaft under grant LE 838/12 and by the COST Action MP1405 QSPACE.

\end{document}